# Superconducting and magneto-transport properties of $BiS_2$ based superconductor $PrO_{1-x}F_xBiS_2$ (x = 0 to 0.9)

Rajveer Jha, Hari Kishan and V.P.S. Awana

CSIR-National Physical Laboratory, Dr. K.S. Krishnan Marg, New Delhi-110012, India

**Abstract**: We report superconducting properties of $PrO_{1-x}F_xBiS_2$ compounds, synthesized by the vacuum encapsulation technique. The synthesized $PrO_{1-x}F_xBiS_2$ (x=0.1, 0.3, 0.5, 0.7 and 0.9) samples are crystallized in a tetragonal P4/nmm space group. Both transport and *DC* magnetic susceptibility measurements showed bulk superconductivity below 4 K. The maximum $T_c$ is obtained for x = 0.7 sample. Under applied magnetic field both $T_c$ onset and $T_c$ ($\rho$ =0) decrease to lower temperatures. We estimated highest upper critical field [$H_{c2}(0)$] for $PrO_{0.3}F_{0.7}BiS_2$ sample to be above 4 T (Tesla). The thermally activated flux flow (*TAFF*) activation energy ($U_0$) is estimated 54.63 meV in 0.05 Tesla field for $PrO_{0.3}F_{0.7}BiS_2$ sample. Hall measurement results showed that electron charge carriers are the dominating ones in these compounds. Thermoelectric effects (Thermal conductivity and Seebeck coefficient) data suggest strong electron-electron correlations in this material.

*Corresponding Author
Dr. V. P. S. Awana, Principal Scientist
E-mail: awana@mail.npindia.org
Ph. +91-11-45609357, Fax-+91-11-45609310
Homepage www.fteewebs.com/vpsawana/

**Introduction:**

Very recently discovered superconductivity in $BiS_2$ based layered compounds has attracted much attention of various groups, because these compounds have similar layered structure as in famous cuprates high $T_c$ superconductors (*HTSc*) and iron Pnictides superconductors. Superconductivity in $Bi_4O_4S_3$ with $T_c$=8.6K and (La/Nd/Ce/Pr)$O_{0.5}F_{0.5}BiS_2$ with $T_c$=3-10 K [1-11] has recently been reported. It is expected that cuprates and iron pnictides compounds, containing 3*d* transition metals, exhibit superconductivity within their layered structures. Superconducting transition temperature can be varied by the changing blocking layer. Here, the basic structural unit, i.e., $BiS_2$ layer is similar to the $CuO_2$ planes in Cu–based superconductors [12] and the FeAs planes in iron pnictides [13]. The parent phase of these compounds i.e., $ReOBiS_2$ (Re= La, Nd, Ce, & Pr) is a bad metals or quasi insulator. The doping of charges carriers improves the electrical conduction and exhibit superconductivity at low temperatures. The doping mechanism of the $BiS_2$ based superconductors is seemingly the same as for Fe pnictides [14, 15]. It has also been suggested that superconductivity emerges in close proximity to an insulating normal state [16]. Substituting F for O induces superconductivity in $BiS_2$-based superconductors. Superconductivity is achieved by doping of the oxide blocks (LaO), which acts as spacer layer. In case of $LaO_{1-x}F_xBiS_2$ superconductivity appears at x = 0.2, reaches a maximum at x = 0.5 and is gradually suppressed and disappeared at x = 0.7 [3, 8, 17]. The superconducting transition temperature is the highest at the optimal doping point of around x = 0.5. Superconductivity has also been observed by electrons doping via substitution of tetravalent $Th^{+4}$, $Hf^{+4}$, $Zr^{+4}$, and $Ti^{+4}$ for trivalent $La^{+3}$ in $LaOBiS_2$ [18]. Superconductivity was also reported in $SrFBiS_2$ compound, via hole doping through substitution of La at the site of Sr [19]. $SrFBiS_2$ seems to be the ground state of these $BiS_2$ based layered superconductors [19]. The pressure dependent resistivity experiments have been performed on $Bi_4O_4S_3$, $LaO_{0.5}F_{0.5}BiS_2$, $NdO_{0.5}F_{0.5}BiS_2$, $CeO_{0.5}F_{0.5}BiS_2$ and $Pr O_{0.5}F_{0.5}BiS_2$ samples [17, 20-22,], and the results indicated the enhancement of superconductivity accompanied by the suppression of semiconducting behavior with increase in pressure. Multiband behavior with dominant electron carriers in these materials has been observed by both experimental and theoretical studies and electron carriers are being known to be originated from the Bi 6p orbital [23]. These newest layered $BiS_2$-based superconducting systems are very sensitive to the carrier doping level, as the atomic substitutions cause profound changes in their properties, and superconductivity appears in the vicinity of the insulating-like state. A chance to explore superconductivity and increase $T_c$ in these new compounds has already resulted in a lot of work that appeared shortly after their discovery [1-11, 18–22]. Hall effect measurements revealed multiband features and suggested that superconducting pairing occurs in one dimensional chains [16]. These compounds do also possess reasonably good thermoelectric properties [24].

In this paper, we report on the synthesis and doping dependence of $T_c$ in $PrO_{1-x}F_xBiS_2$ compounds. All samples are crystallized in tetragonal *P4/nmm* space group, the *c* lattice parameter decrease with the doping of F for x=0.7 and then slightly increase for higher concentration. Superconducting temperature ($T_c^{onset} \sim 4.5$ K) is highest for $PrO_{0.3}F_{0.7}BiS_2$ sample. Here we present structural, electrical and thermal transport properties of the PrO$_{1-x}$F$_x$BiS$_2$ samples.

**Experimental:**

Bulk polycrystalline PrO$_{1-x}$F$_x$BiS$_2$ (x=0.1, 0.3, 0.5, 0.7 & 0.9) samples were synthesized by standard solid state reaction route via vacuum encapsulation. High purity Pr, Bi, S, PrF$_3$, and Pr$_6$O$_{11}$ are weighed in stoichiometric ratio and ground thoroughly in a glove box under high purity argon atmosphere. The mixed powders are subsequently palletized and vacuum-sealed (10$^{-3}$ Torr) in a quartz tube. Sealed quartz ampoules were placed in box furnace and heat treated at 780$^0$C for 12h with the typical heating rate of 2$^o$C/min., and subsequently cooled down slowly to room temperature. This process was repeated twice. X-ray diffraction (*XRD*) was performed at room temperature in the scattering angular (*2θ*) range of 10$^o$-80$^o$ in equal *2θ* step of 0.02$^o$ using *Rigaku diffractometer* with *Cu K$_α$* ($\lambda$ = 1.54Å). Rietveld analysis was performed using the standard *FullProf* program. Detailed dc and ac transport and magnetization measurements were performed on Physical Property Measurements System (*PPMS*-14T, *Quantum Design*).

**Results and discussion:**

Figure 1 shows the room temperature observed and Reitveld fitted XRD pattern of as synthesized PrO$_{1-x}$F$_x$BiS$_2$ (x = 0.1 to 0.9) samples. The compounds are crystallized in tetragonal structure with space group *P4/nmm*. Small amount of Bi and Bi$_2$S$_3$ impurities are also seen within main phase of the compound. As seen from the XRD patterns in Figure 1, the impurity phases are quite small and present in all the samples. Though, we did not quantify exactly the amount of impurities present in main phase of the samples, but the same does not appear to have direct impact on observation of superconductivity. The Rietveld fitted results show that the *a* axis of PrO$_{1-x}$F$_x$BiS$_2$ is 4.014(2) Å for *x* = 0.1, which increases to 4.019(1) Å for *x* = 0.7, and then slightly decreases to 4.018(2) Å for *x* = 0.9. The *c*-axis lattice constant decreases from 13.508(3) Å to 13.360(1) Å as *x* is increased from *x* = 0.1 to 0.7 and the same increase slightly to *c* = 13.386(4) Å for x = 0.9. This result indicates that the layer structure slightly expands in the in-plane direction and compressed in *c*-direction with F doping, reaches a maximum at *x* = 0.7, and then starts to shrink at *x* = 0.9. Decrease of the *c*-axis lattice parameter indicate that F has been successfully substituted at the O site, as the ionic radius of F is smaller than that of O. Similar trend of lattice parameters is reported for NdO$_{1-x}$F$_x$BiS$_2$ and CeO$_{1-x}$F$_x$BiS$_2$ compounds [9, 16]. As far as increase of the *c*-axis lattice parameter for x = 0.9 sample is concerned, we believe that the doping of F at O site in PrOBiS$_2$ is successful only up to x = 0.7.

DC magnetic susceptibility under the applied magnetic field of 10 Oe is shown in Figure 2. Magnetization measurements are performed in both ZFC (Zero Field Cooled) and FC (Field Cooled) protocols for the studied PrO$_{1-x}$F$_x$BiS$_2$ (x=0.3, 0.5, 0.7, and 0.9) samples. The compounds exhibit diamagnetic transition at around 2.5 K for x = 0.3, 3.0 K for x = 0.5, 3.7 K for x = 0.7, and at 3.5 K for x = 0.9. A superconducting transition is observed for all the F-doped samples. All samples show reasonable shielding volume fraction, indicating the appearance of bulk superconductivity in these samples. Namely, the same is around 8%, 43%, 80% and 33% for x = 0.3, 0.5, 0.7 and 0.9 samples of PrO$_{1-x}$F$_x$BiS$_2$ series. The PrO$_{0.3}$F$_{0.7}$BiS$_2$ sample exhibits the maximum shielding volume fraction of ~ 80% and the highest T$_c$ (3.7 K) as well, among all the studied samples. Inset of Figure 2, shows the ac magnetic susceptibility of PrO$_{0.3}$F$_{0.7}$BiS$_2$ sample, confirming the bulk superconductivity at around 3.7 K. The ac magnetic susceptibility measurements have been carried out at varying amplitude from 3-10 Oe and fixed frequency 333 Hz. It can be seen that with change in amplitude from 3-10 Oe the imaginary part peak height is increased along with increased diamagnetism in real part of ac susceptibility. The interesting part is that with increasing AC amplitude the imaginary part peak position temperature (3.7 K) is not changed. This is an indication that the superconducting grains are well coupled PrO$_{0.3}$F$_{0.7}$BiS$_2$ superconductor.

The temperature dependence of the resistance for the PrO$_{1-x}$F$_x$BiS$_2$ (x = 0, 0.1, 0.3, 0.5, 0.7, and 0.9) samples is shown in Figure 3. Pure PrOBiS$_2$ sample shows semiconducting behavior at low temperatures without appearance of superconductivity down to 2 K. With doping of F at O site, superconductivity appears and $T_c$ is seen increasing till x = 0.7 further the same is decreased for x = 0.9. A superconducting transition is observed for all the F doped samples. The samples with x =0.7 exhibit high $T_c$ onset of above 4.5 K, inset of the figure shows the zoom part of the *R-T* plots from 1.5- 5 K. The superconducting transition temperature ($T_c$) is slightly higher as being seen from the magnetization measurements (Figure. 2) than the magnetization studies (Fig. 2). This may be due to the fact that the transport measurements require only the percolation path for the resistance less current to flow. On the other hand the magnetization measurements require reasonable diamagnetic fraction for the superconductivity to be seen. However, the general trend of the variation of $T_c$ with x is the same as being seen from both transport and magnetization measurements. The x dependence of lattice parameter *c* (Å) and $T_c$ (magnetization) is plotted in Figure 4. The $T_c$ of the all samples varies with the lattice constant as well as x concentration. It is highest for lowest $c$ lattice constant. This indicates that the chemical pressure is increasing with the F concentration till x = 0.7 and at further concentration of x= 0.9, the $T_c$ is decreased. It is concluded that with decreasing lattice constant the superconducting transition $T_c$ is increasing. As mentioned in XRD results, the doping of F at O site in PrOBiS$_2$ is seemingly successful only up to x = 0.7, perhaps this is the reason that $T_c$ is not increased beyond x = 0.7.

The temperature dependence of resistivity under applied magnetic field is shown in Figure 5 (a), (b), and (c) for the three samples i.e., x = 0.5, 0.7, and 0.9 respectively. $T_c^{onset}$ decreases less compared to $T_c(R=0)$ with the increasing magnetic field, hence the transition width is broadened. Similar broadening of the transition was observed in the high-$T_c$ layered cuprate and Fe-pnictide superconductors. The upper critical field $H_{c2}$ versus $T$ for PrO$_{0.5}$F$_{0.5}$BiS$_2$, PrO$_{0.3}$F$_{0.7}$BiS$_2$ and PrO$_{0.1}$F$_{0.9}$BiS$_2$ is shown in the Figure 6. $H_{c2}$ is estimated using the conventional one-band Werthamer–Helfand–Hohenberg (*WHH*) equation, i.e., $H_{c2}(0) = -0.693T_c(dH_{c2}/dT)_{T=Tc}$. The $H_{c2}$ corresponds to the temperatures, where the resistivity drops to 90% of the normal state resistivity $\rho_n(T,H)$ at $T_c^{onset}$ in applied magnetic fields. The solid lines are the result of fitting of $H_{c2}(T)$ to the *WHH* formula. The estimated $H_{c2}(0)$ is 2.7 Tesla for PrO$_{0.5}$F$_{0.5}$BiS$_2$, 4.8 Tesla for PrO$_{0.3}$F$_{0.7}$BiS$_2$, and 3.8 Tesla for PrO$_{0.1}$F$_{0.9}$BiS$_2$.

The temperature derivatives of resistivity (d$\rho$/d$T$) for the superconducting samples PrO$_{0.5}$F$_{0.5}$BiS$_2$, PrO$_{0.3}$F$_{0.7}$BiS$_2$ and PrO$_{0.1}$F$_{0.9}$BiS$_2$ are shown in Figure.7 (a), (b) and (c) respectively. It is clear that only a single transition peak is seen for every applied field. This single superconducting transition suggests better grains coupling in these systems. Also we observed some broadening of peaks with increasing applied magnetic field. The broadening of d$\rho$/d$T$ peak for increasing applied field suggests that superconducting onset is relatively less affected than the $T_c$ ($\rho$=0) state. Due to the thermally activated flux flow (*TAFF*) there is broadening of resistive transitions under applied magnetic fields [25]. The resistance in broadened region is due to the creep of vortices, which are thermally activated. The resistivity in this region can be given by Arrhenius equation [26, 27],

$$\rho(T,B) = \rho_0 \exp[ - U_0/k_BT]$$

Where, $\rho_0$ is the field independent pre-exponential factor (the normal state resistance at 5K i.e., $\rho_{5K}$ is taken as $\rho_0$), k$_B$ is the Boltzmann's constant and $U_0$ is *TAFF* (thermally activated flux flow) activation energy, which can be obtained from the slope of the linear part of an Arrhenius plot in low resistivity region. We have plotted experimental data as ln($\rho/\rho_{5K}$) versus $T^{-1}$ in Figure 8 (a), (b) and (c) for all three samples. The best fitted data give values of the activation energy, which varies from 54.63 meV to 1 meV in the magnetic field range of 0.05 to 1 Tesla for the $PrO_{0.3}F_{0.7}BiS_2$ sample. The magnetic field dependence of activation energy ($U_0$) for $PrO_{0.5}F_{0.5}BiS_2$, $PrO_{0.3}F_{0.7}BiS_2$ and $PrO_{0.1}F_{0.9}BiS_2$ samples is shown in Figure 9.

Hall effect measurements of superconducting $PrO_{0.3}F_{0.7}BiS_2$ sample are shown in the figure 10. The temperature dependence of $R_H$ has been carried out at a magnetic field of 1 Tesla from 300 K down to 2 K. The negative $R_H$ signal is observed for the given temperature range, indicating that the dominant charge carriers are electrons similar to that as in iron based superconductors [19]. The inset of the Figure 10, depicts magnetic field dependent $\rho_{xy}$ at different temperatures from 2 – 300 K of the same sample. Hall resistivity $\rho_{xy}$ is measured with longitudinal current and perpendicular magnetic field ($H$ = 0-5 Tesla) to the surface of the sample and the voltage $V_{xy}$ is measured across the direction of sample width. Here $\rho_{xy}$ remains negative for all temperatures, indicating that the electrons are the dominating current carriers in the system. Further the $\rho_{xy}$ exhibited curved transition like feature at 2 K, below say 1.5 Tesla. For higher temperatures (10 K – 300 K) the resistivity remains negative and linear. The curved feature below 1.5 Tesla at 2K happens due to the fact that the compound is superconducting below this temperature and applied fields. For higher fields (> 1.5 Tesla) the $V_{xy}$ at 2 K is again linear and negative because the compound is turned to normal conductor at this field. When $\rho_{xy}$ exhibits a linear behavior with the magnetic field, the $R_H$ can be measured by $R_H = d\rho_{xy}/dH = 1/ne$, for a single band metal, where $n$ is the charge carrier density. Using a single band assumption and taking the value of $R_H$ at 5 K, we get the charge carrier density of $3.84 \times 10^{19}/cm^3$. Worth caution is the fact that the single band metal model could not work for the $BiS_2$ based superconductors [16] and the situation may be similar to the case of Fe pnictides and hence the carrier calculation may need corrections.

Figure 11 shows thermal conductivity κ ($T$) and Seebeck coefficient $S(T)$ results of $PrO_{0.3}F_{0.7}BiS_2$ sample. Thermal conductivity in the $PrO_{0.3}F_{0.7}BiS_2$ compound is seemingly dominated by phonons and the electron contribution being calculated by the Wiedemann–Franz law may be negligible. Similar situation is found for $SmFeAs(O_{0.93}F_{0.07})$ [28]. Smaller κ values in $PrO_{0.3}F_{0.7}BiS_2$ may represent the crystallographic disorder. An enhancement of κ value above the superconducting temperature 5 K is due to the gap opening at the Fermi surface, followed by carrier condensation and consequent suppression of electron-phonon scattering. Seebeck coefficient $S(T)$ is linearly dependent with temperature. The negative $S(T)$ for whole measurement temperature region, indicates the electron-like carriers. The $S(T)$ results support the Hall measurements data. Relatively larger $S$ of around 30μV/K and low thermal conductivity of 1W/mK is reminiscent of strong electron correlation in these systems. Very recently the single crystals [29, 30] of ReOBiS$_2$ (Re= La, Nd, Ce, & Pr) are being grown successfully and their thermal transport data could presumably lend more authentic information on these newest class of superconductors.

**Conclusion:**

In conclusion, the synthesized BiS$_2$-based superconductor PrO$_{1-x}$F$_x$BiS$_2$ (x=0.1-0.9) is crystallized in a tetragonal *P*4/*nmm* space group. Reitveld refinement results indicated successful substitution of F at O site. Both electrical resistivity and *DC* magnetic susceptibility measurements showed bulk superconductivity below 4K. The upper critical field [$H_{c2}(0)$] for the optimized PrO$_{0.3}$F$_{0.7}$BiS$_2$ sample is estimated over 4 Tesla from $\rho(T,H)$ measurements. The flux flow activation energy is estimated to be 54.63 meV in 0.05 Tesla field for PrO$_{0.3}$F$_{0.7}$BiS$_2$ sample. Hall measurement and thermal transport results indicate dominance of electron charge carriers along with strong electron correlation in this compound. Though the $T_c$ of F doped ReOBiS$_2$ is relatively low (2-10K), yet their layered structure, the character ground state, doping pattern, broadening of superconducting transition under magnetic field along with higher S and low κ could put them close to high $T_c$ cuprates and Fe pinctide superconductors.

**Acknowledgement:**

Authors would like to thank their Director Professor R.C. Budhani for his keen interest in the present work. This work is supported by *DAE-SRC* outstanding investigator award scheme on search for new superconductors. Rajveer Jha acknowledges the *CSIR* for the senior research fellowship. H. Kishan thanks CSIR for providing Emeritus Scientist Fellowship.

**Figure Captions**

**Figure 1:** Observed (*open circles*) and calculated (*solid lines*) XRD patterns of $PrO_{1-x}F_xBiS_2$ (x=0.1, 0.3, 0.5, 0.7 and 0.9) compound at room temperature.

**Figure 2:** *DC* magnetization (both *ZFC* and *FC)* plots for $PrO_{1-x}F_xBiS_2$ (x=0.3, 0.5, 0.7 and 0.9) measured in the applied magnetic field, $H$= 10 Oe. Inset show the ac magnetic susceptibility in real (*M'*) and imaginary(*M''*) situations at fixed frequency of 333 Hz in varying amplitudes of 3–10 Oe for $PrO_{0.3}F_{0.7}BiS_2$ sample.

**Figure 3:** Resistance versus temperature (*$R/R_{300}$* Vs *T*) plots for $PrO_{1-x}F_xBiS_2$ (x=0, 0.1, 0.3, 0.5, 0.7 & 0.9) samples, inset show the same in 1.5-5.0 K temperature range.

**Figure 4:** The concentration of F-doping (x) dependence of the lattice constants (*c*-axis) and superconducting $T_c$.

**Figure 5:** Temperature dependence of the resistivity $\rho(T)$ under magnetic fields for the (a) $PrO_{0.5}F_{0.5}BiS_2$, (b) $PrO_{0.3}F_{0.7}BiS_2$ and (c) $PrO_{0.1}F_{0.9}BiS_2$ samples.

**Figure 6:** The upper critical field $H_{c2}$ taken from 90% of the resistivity $\rho(T)$ for the samples $PrO_{0.5}F_{0.5}BiS_2$, $PrO_{0.3}F_{0.7}BiS_2$ and $PrO_{0.1}F_{0.9}BiS_2$.

**Figure 7:** Temperature derivative of normalized resistivity Versus *T* for (a) $PrO_{0.5}F_{0.5}BiS_2$, (b) $PrO_{0.3}F_{0.7}BiS_2$ and (c) $PrO_{0.1}F_{0.9}BiS_2$ samples.

**Figure 8:** Fitted Arrhenius plots of resistivity for $PrO_{0.5}F_{0.5}BiS_2$, $PrO_{0.3}F_{0.7}BiS_2$ and $PrO_{0.1}F_{0.9}BiS_2$ samples.

**Figure 9:** Magnetic field dependent *TAFF* activation energy plots on log-log scale, the solid lines represent power law fitting to the experimental data of all three samples.

**Figure 10:** The Hall coefficient $R_H$ of the $PrO_{0.3}F_{0.7}BiS_2$ sample at 1 Tesla from 2 to 300 K. inset show the Hall resistivity $\rho_{xy}$ vs the magnetic field $\mu_0 H$ at 2, 10, 50, 100, 200, and 300 K for the same sample.

**Figure 11:** Temperature dependence of Thermal conductivity κ (*T*) and Seebeck coefficient *S* (*T*) for the sample $PrO_{0.3}F_{0.7}BiS_2$.

Fig. 1

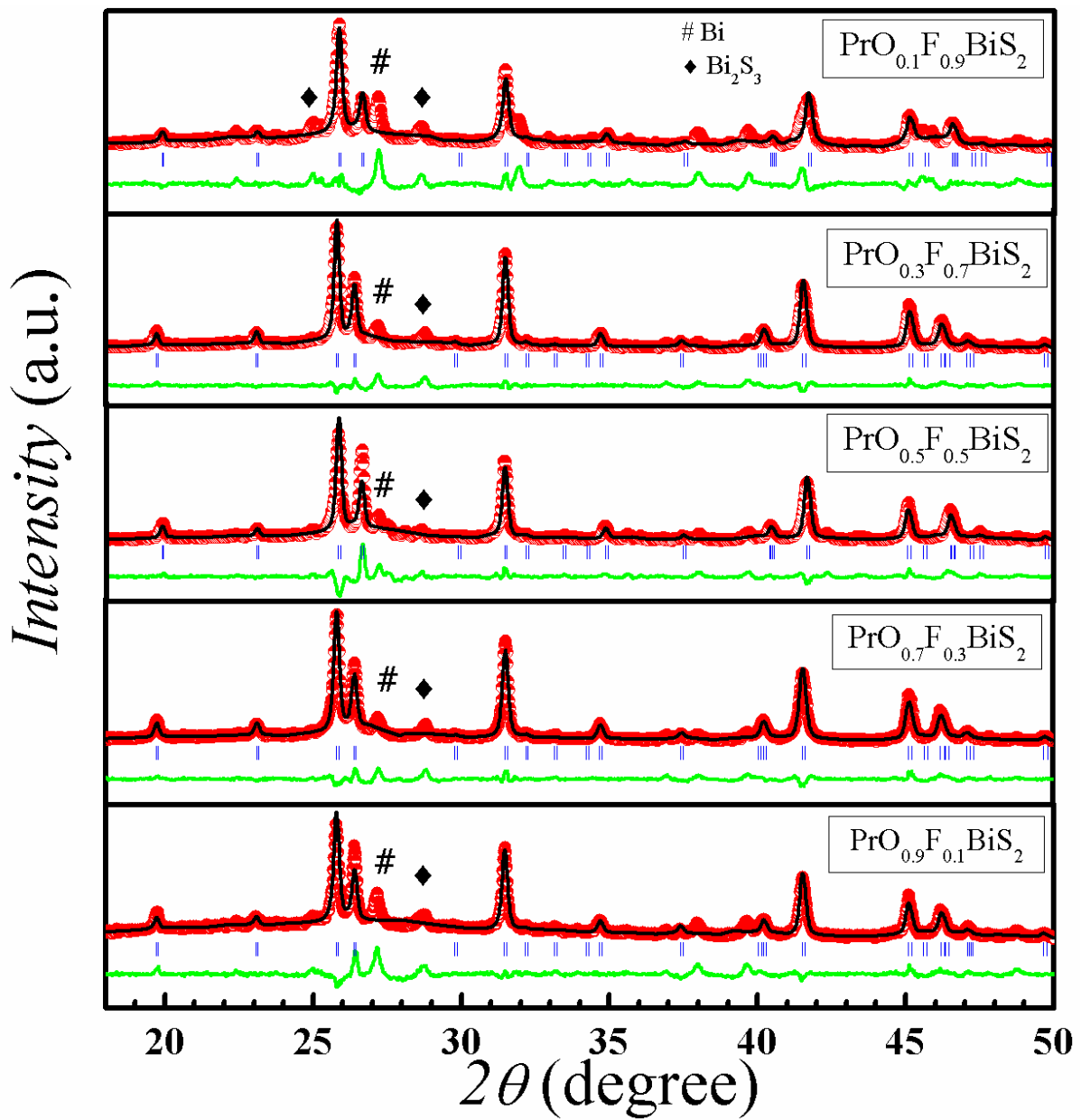

Fig. 2

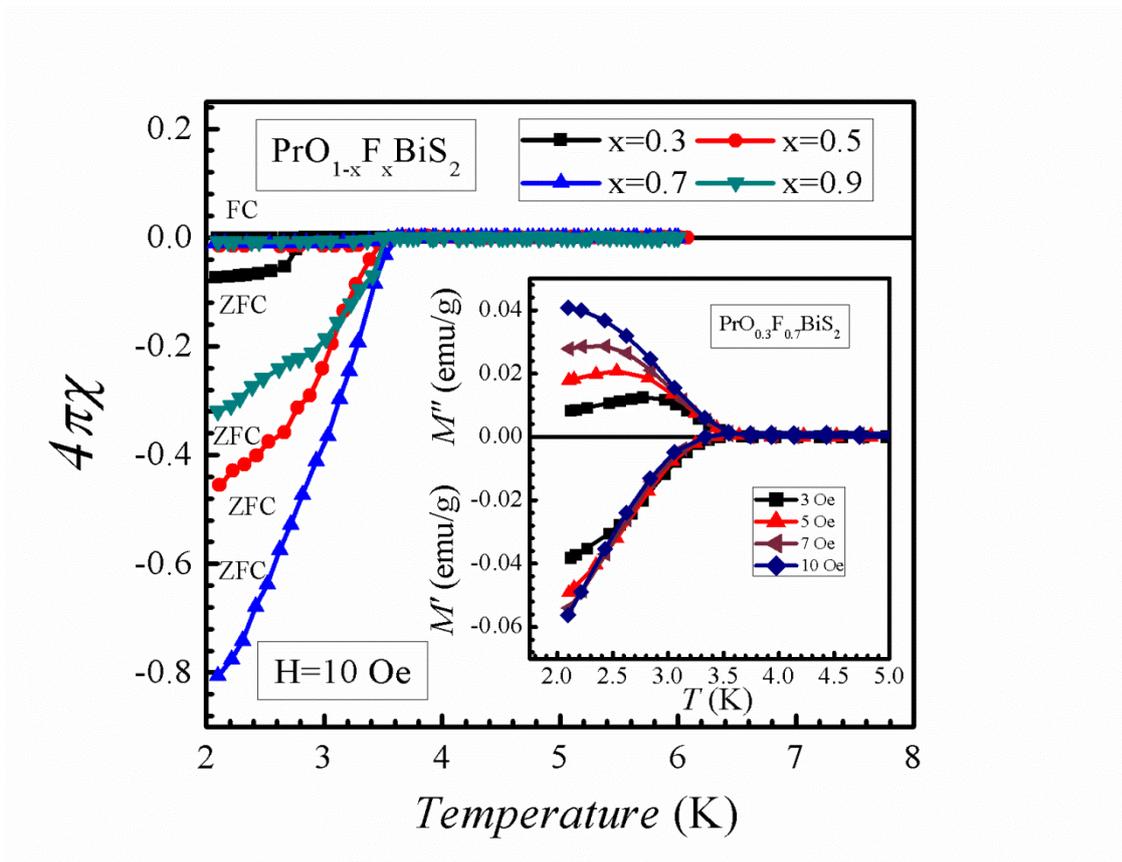

Fig.3

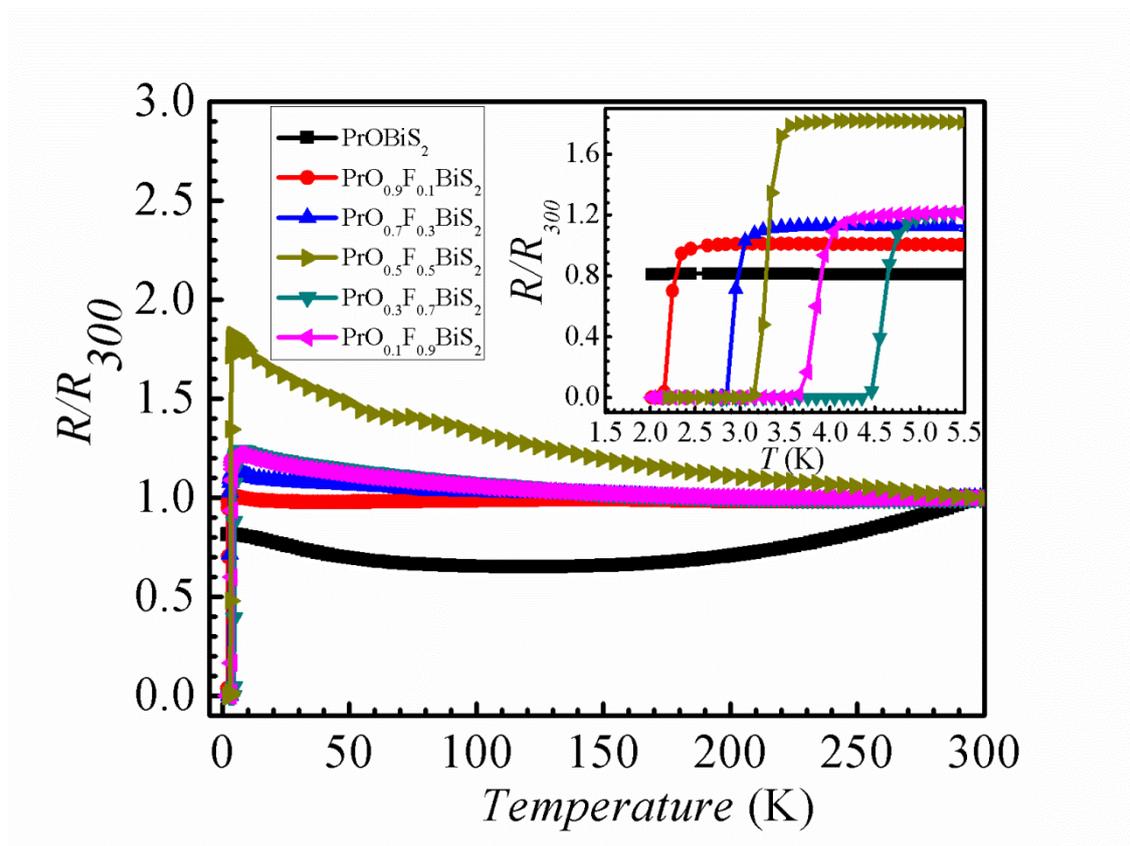

Fig.4

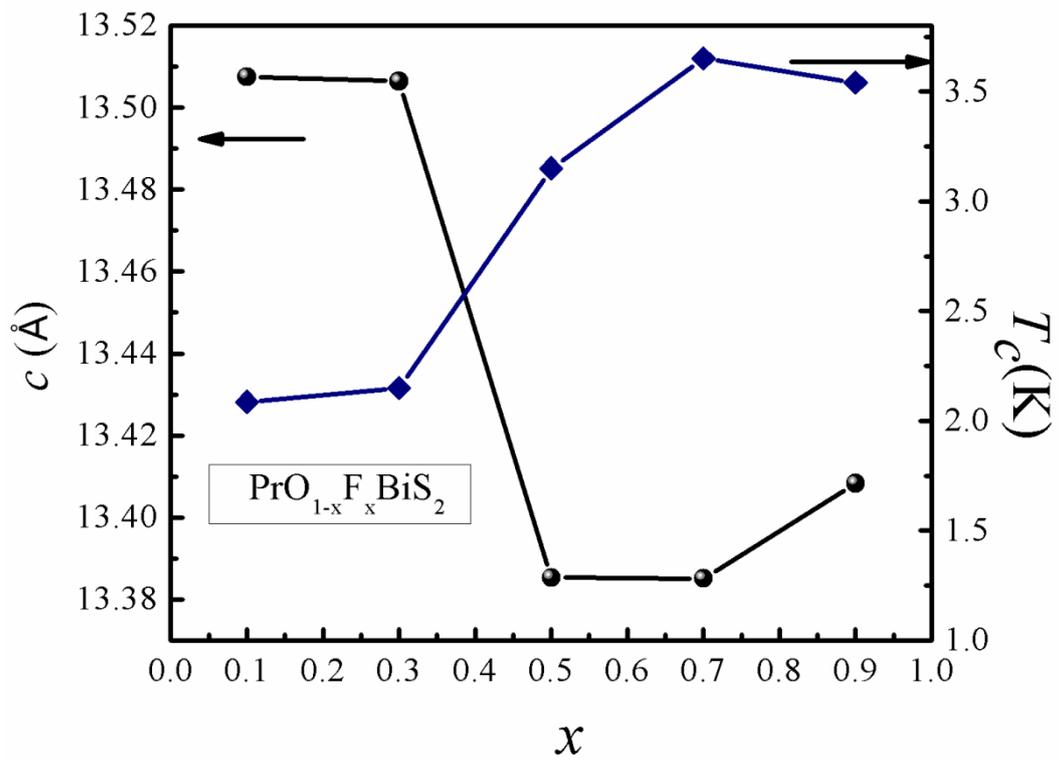

Fig.5

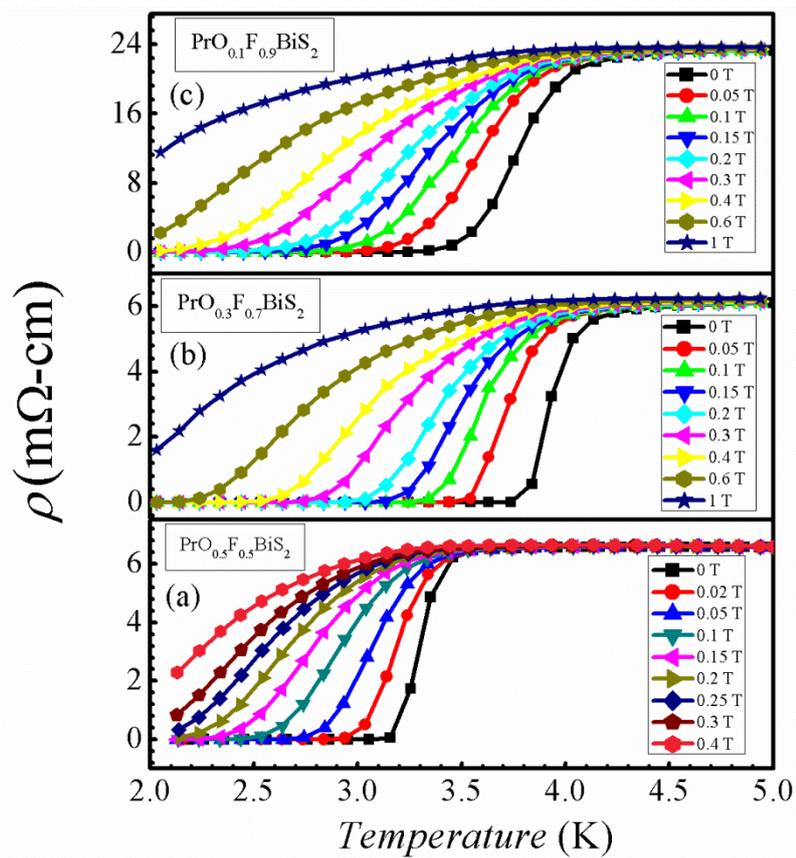

Fig. 6

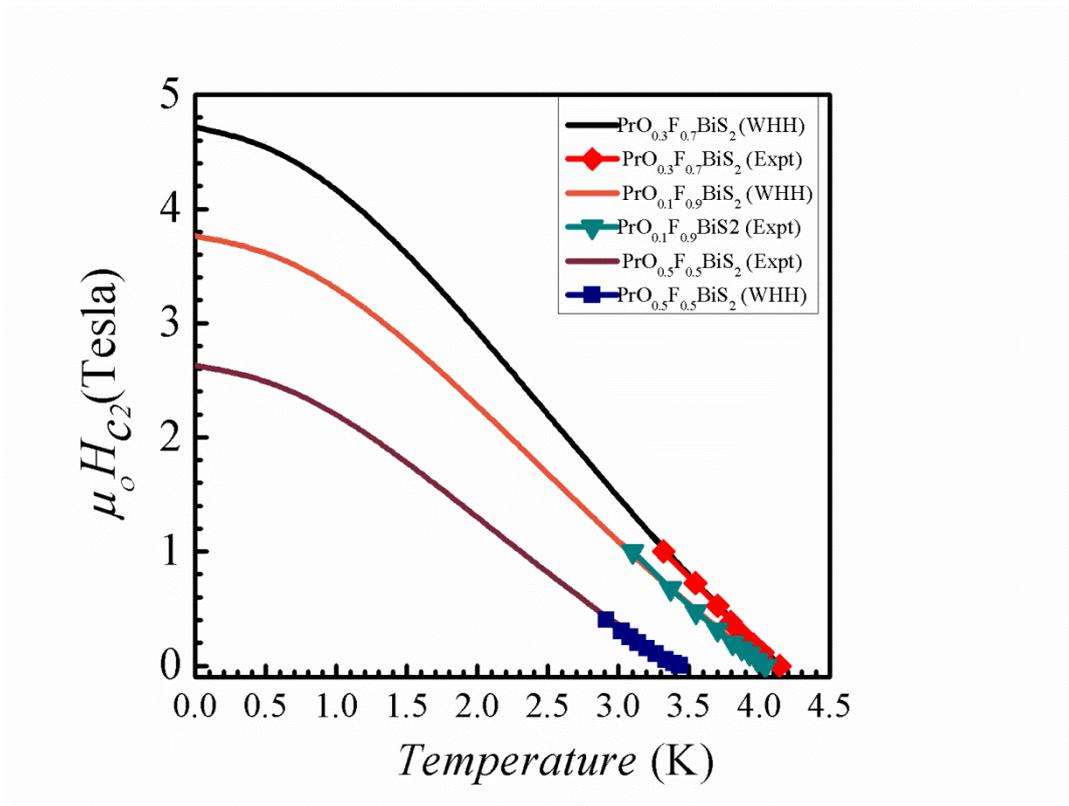

Fig.7

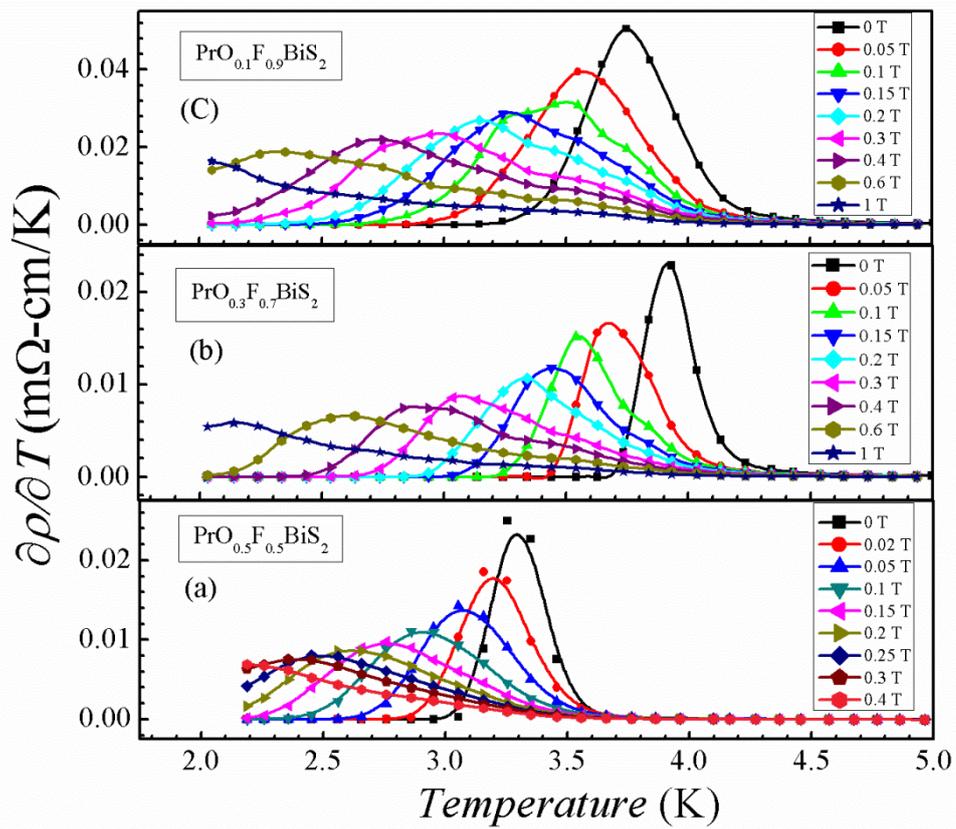

Fig.8

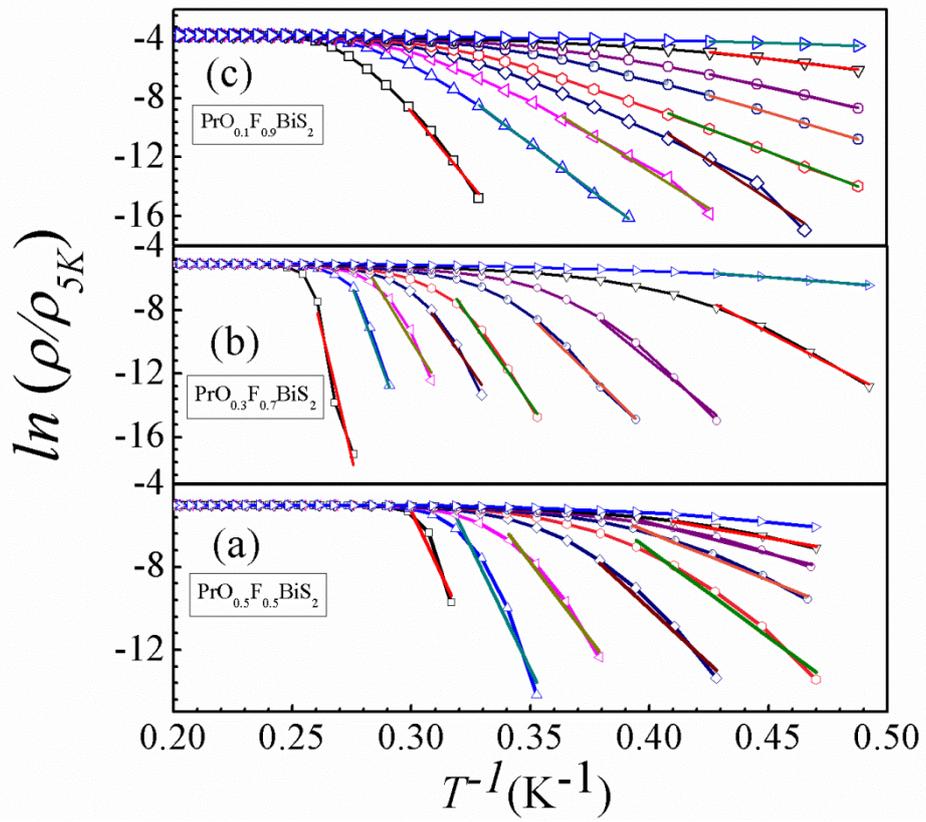

Fig.9

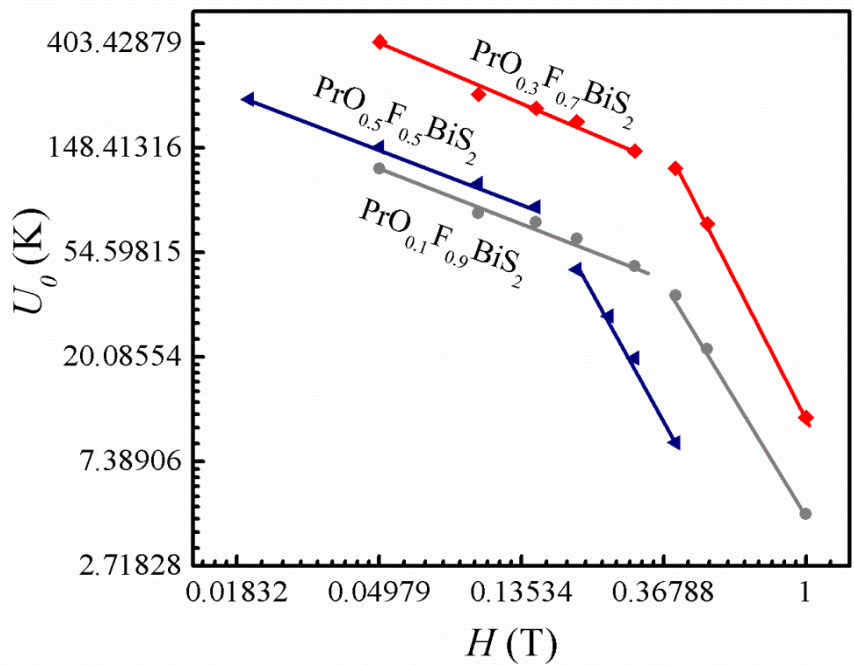

Fig.10

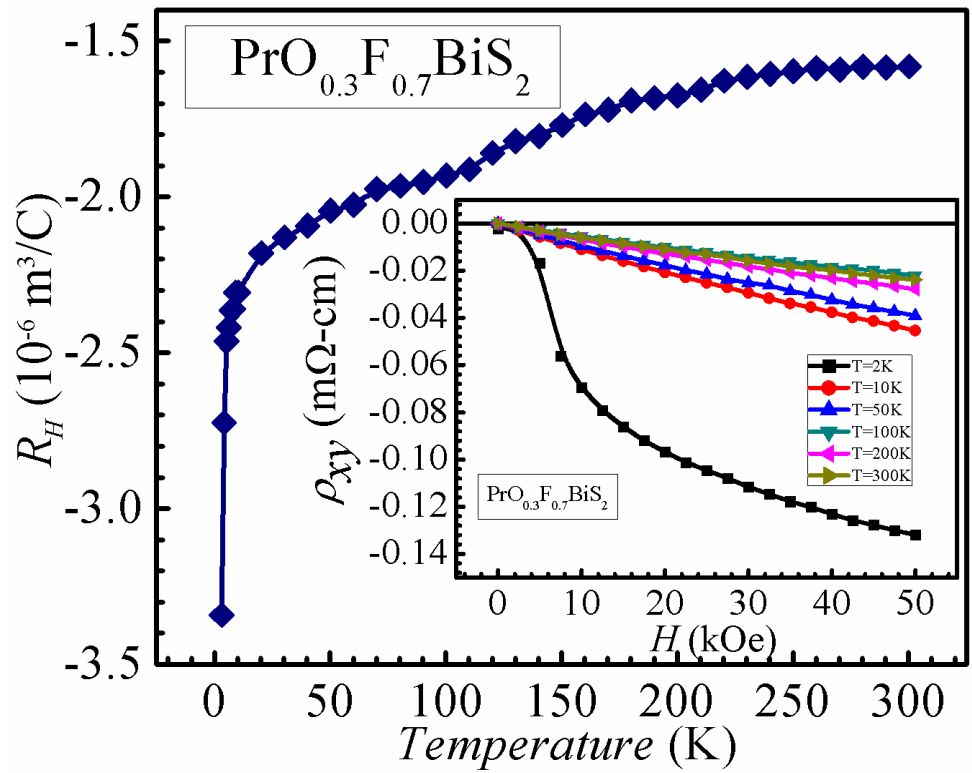

Fig.11

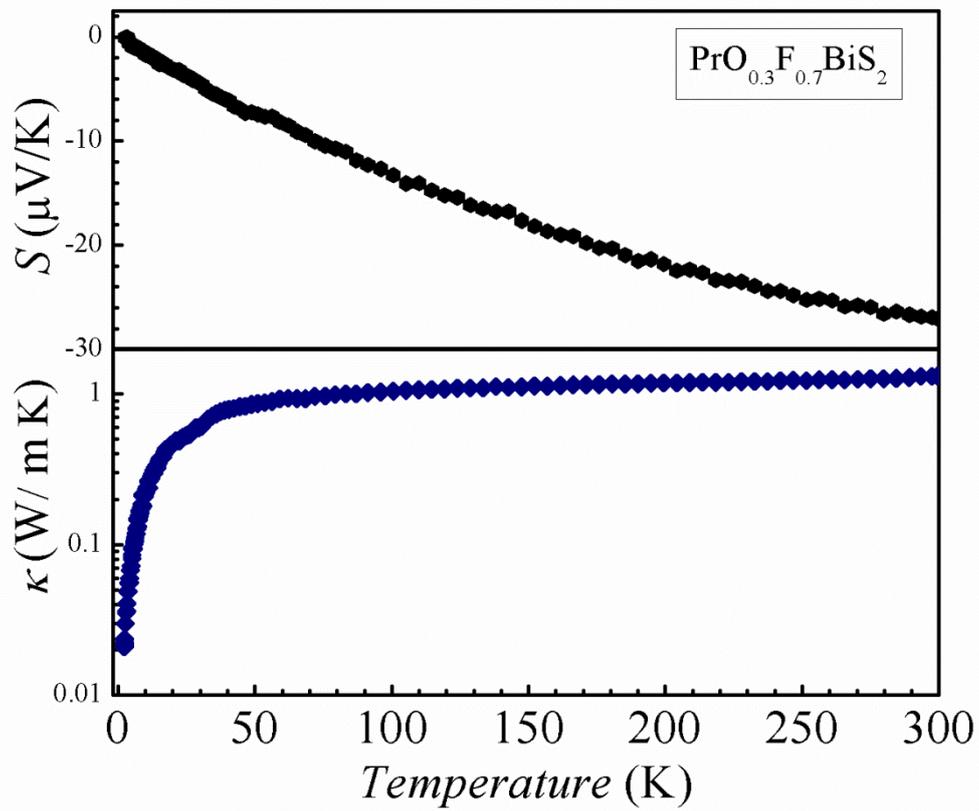